\documentclass[a4paper,12pt]{article}
\usepackage[english]{babel}
\usepackage{graphicx}

\input epsf.tex

\begin{document}

\centerline {\bf Preparation of quantum states of two spin-$\frac{1}{2}$ particles}
\centerline {\bf in the form of the Schmidt decomposition}
\medskip
\centerline {A. R. Kuzmak$^1$, V. M. Tkachuk$^2$}
\medskip
\centerline {\small \it Department for Theoretical Physics, Ivan Franko National University of Lviv,}
\medskip
\centerline {\small \it 12 Drahomanov St., Lviv, UA-79005, Ukraine}
\medskip
\centerline {\small \it E-Mail: $^1$andrijkuzmak@gmail.com, $^2$voltkachuk@gmail.com}
\medskip

{\small

We consider a system of two spins that are coupled via an isotropic Heisenberg Hamiltonian. For the first time, a two-step method for the preparation
of an arbitrary quantum state of two qubits in the form of the Schmidt decomposition is proposed. The simplified version of this method is applied to the
physical system of an atom having with a nuclear spin $1/2$ and one valence electron. As an example, the preparation of two-spin quantum
states in the $^{31}$P system is considered.
\medskip

}

\section{Introduction\label{sec1}}

Physical implementation of quantum computation requires systems isolated from their environment, for providing a high degree of quantum
coherence \cite{quantcomp}. States of these systems must be measured with high fidelity.
Physical systems, which have high rate of quantum coherence and that can be measured with high accuracy, were suggested in many papers:
electron spins in quantum dots \cite{qdots1, qdots2, qdots3}, electron spins
in a semiconductor \cite{qdots2, scc, semcond1, semcond2, semcond3, semcond4, semcond5, semcond6}, superconducting qubits
\cite{supcond1, supcond2, supcond3, supcond4} and the $^{31}$P donor in silicon
\cite{phosphorus3, phosphorus1, phosphorus2, phosphorus4, phosphorus6, phosphorus7, phosphorus8}.

Manipulation of spin particles is essential for spin-based quantum calculations. The spin resonance techniques \cite{srtech} for coherent control of
spin particles \cite{scc} is well-developed, therefore wide attention has been attracted to spin qubits \cite{sq1, sq2, sq3, sq4}. Fast manipulation and
long-lived coherence of spin particles make electron-nuclear spin systems promising candidates for quantum computing.

There is growing interest in spin systems with the electron-nuclear interaction, such as bismuth in silicon \cite{BiinSi} or hydrogen in silicon
\cite{HinSi}, but the most popular system is a phosphorus donor in silicon \cite{phosphorus3}.
Interest in quantum computation on a phosphorus donor in silicon has increased after publication of the paper \cite{phosphorus3}. In this article a scheme
for implementation of a silicon-based nuclear spin quantum computer was presented. The $^{31}$P has one nuclear spin ($I=1/2$)
and one electron spin ($S=1/2$). Therefore, it can be considered as a two-qubit system \cite{phosphorus3}.
Long coherence time of the electron \cite{phosphorus6} and of the nuclear \cite{phosphorus8, LNT} spin of the $^{31}$P donor in silicon makes them
promising building blocks for the realization of a solid-state quantum computer.
Electrical detection and a coherent manipulation of the $^{31}$P electronic states \cite{phosphorus2, phosphorus9, phosphorus10, phosphorus11, phosphorus12}
and the nuclear states \cite{phosphorus1, phosphorus7, phosphorus8} have been shown.

In \cite{CRSP} several new protocols for the controlled remote state preparation (CRSP) by using a five-particle Brown state as a quantum channel were
proposed. Primarily, they proposed CRSP protocol of arbitrary two and three qubit states. The CRSP of arbitrary two qubit state is completed by
Alice under the permission of Supervisor. Alisa wants to prepare an arbitrary two qubit state to the remote receiver Bob. Supervisor does not know the
details of the initial state but decides whether the task should be completed or not.
A similar scheme for the remote preparation of arbitrary two-qubit entangled state, where two GHZ states are considered as the
quantum channel, was proposed in \cite{GHZP}.

In this paper, we suggest for the first time a two-step method for the preparation of arbitrary quantum state of two qubits (Sec. \ref{sec2}). This method
naturally allows to prepare a predefined quantum state in the form of the Schmidt decomposition. At the first step the quantum state
driven by an isotropic Heisenberg Hamiltonian, evolves from an initial state to some state which depends on the period of time of the evolution $t_1$ (Sec. \ref{sec2}).
At the second step, at the moment of time $t_1$ we apply pulsed magnetic fields individually to each spin and obtain the final quantum state (Sec. \ref{sec2}).
A simplified version of this method is applied to a physical system of an atom which has a nuclear spin $1/2$ and a
valence electron (Sec. \ref{sec3}). Conditions necessary for the preparation of some two-qubit quantum state in the $^{31}$P system are represented in Sec. \ref{sec4}.
In Sec. \ref{sec5} we give conclusions.

\section{A two-step preparation of quantum states in the form of the Schmidt decomposition\label{sec2}}

In this section the method for preparing arbitrary quantum state of two spins with an isotropic Heisenberg Hamiltonian is proposed.
The method we suggest consists of two steps. At the first step, by using the evolution operator with an isotropic
Heisenberg Hamiltonian the state, which depends on the period of time of evolution $t_1$, is obtained. At the second step, at the moment of time $t_1$ we apply
pulsed magnetic field individually to each spin and obtain the final state. Let us consider each step in details.

{\it Step I}. We assume that a two-spin physical system interacts via an isotropic Heisenberg Hamiltonian with coupling $A$
\begin{eqnarray}
H_{I}=\frac{A}{4}\left(\sum_{i=x,y,z}\sigma_i^1\sigma_i^2+1\right),
\label{eq1}
\end{eqnarray}
where $\sigma_i^1=\sigma_i\otimes 1$, $\sigma_i^2=1\otimes\sigma_i$ and $\sigma_i$ are the Pauli matrices.
The Hamiltonian (\ref{eq1}) has one three-fold degenerate eigenvalue $\frac{A}{2}$ (triplet state)
\begin{eqnarray}
&&\vert T_+\rangle =\vert\uparrow\uparrow\rangle,\label{eq2(11)}\\
&&\vert T_-\rangle =\vert\downarrow\downarrow\rangle,\label{eq2(12)}\\
&&\vert T_0\rangle =\frac{1}{\sqrt{2}}\left(\vert\uparrow\downarrow\rangle + \vert\downarrow\uparrow\rangle\right),
\label{eq2(1)}
\end{eqnarray}
and eigenvalue $\left(-\frac{A}{2}\right)$ with singlet state
\begin{eqnarray}
\vert S\rangle =\frac{1}{\sqrt{2}}\left(\vert\uparrow\downarrow\rangle - \vert\downarrow\uparrow\rangle\right).
\label{eq2(2)}
\end{eqnarray}
The evolution operator for this system takes the form
\begin{eqnarray}
U_I=e^{-iH_{I}t}=\cos\left(\frac{A}{2}t\right)-i\frac{2}{A}\sin\left(\frac{A}{2}t\right)H_I.
\label{eq3}
\end{eqnarray}
Here we use the fact that $H_{I}^2=\left(\frac{A}{2}\right)^2$. We set $\hbar =1$ and it means that the coupling parameter $A$ is measured in frequency units.
Note that the Hamiltonian (\ref{eq1}) is a special case of the Hamiltonian considered in our previous paper where the brachistochrone problem
for two spin particles was examined \cite{brach}.

Let us consider evolution of the system of two spins having started from the initial state $\vert\uparrow\downarrow\rangle$. The initial state
can be easily created because this state is the eigenstate of the system of two spins in strong magnetic field, which is directed along the $z$-axis.
Note that we do not start from $\vert\uparrow\uparrow\rangle$ or $\vert\downarrow\downarrow\rangle$ states
because these states are eigenvectors of the Heisenberg Hamiltonian (\ref{eq1}). Hence, the state, which is a result of the interaction between two spins during
period of time $t_1$, can be represented in the form
\begin{eqnarray}
\vert\psi_I\rangle =e^{-iH_{I}t_1}\vert\uparrow\downarrow\rangle=\cos\left(\frac{A}{2}t_1\right)\vert\uparrow\downarrow\rangle + \sin\left(\frac{A}{2}t_1\right)e^{-i\frac{\pi}{2}}\vert\downarrow\uparrow\rangle.
\label{eq4}
\end{eqnarray}

{\it Step II}. The state (\ref{eq4}) is defined by one real parameter $t_1$. An arbitrary quantum state of two qubits contains six real parameters. Due to
this fact it is reasonable to apply the pulsed magnetic fields individually to each spin
\begin{eqnarray}
H_{II}=\chi_1\mbox{\boldmath{$ \sigma$}}^1 \cdot {\bf n}^1\delta(t-t_1)+\chi_2\mbox{\boldmath{$ \sigma$}}^2 \cdot {\bf n}^2\delta(t-t_1),
\label{eq5}
\end{eqnarray}
where $\delta(t-t_1)$ is Dirac's delta function, $\bf{n}^{\it i}=\left(\sin\theta_{\it i}\cos\phi_{\it i},\sin\theta_{\it i}\sin\phi_{\it i},\cos\theta_{\it i}\right)$,
$\phi_i$ and $\theta_i$ determine the direction of the magnetic field for the first ($i=1$) and the second ($i=2$) spins, respectively.
Dirac's delta function allows us to neglect the interaction between spins when the magnetic field is applied. Indeed, let us consider
the operator of evolution for the system of two interacting spins, which is controlled by the external magnetic fields (\ref{eq5}) during
a sufficiently short period of time $\tau$
\begin{eqnarray}
U_{II}=\exp{\left[-i\int_{t_1}^{t_1+\tau}\left(\frac{A}{4}\left(\sum_{i=x,y,z}\sigma_i^1\sigma_i^2+1\right)+\left(\chi_1\mbox{\boldmath{$ \sigma$}}^1 \cdot {\bf n}^1 +\chi_2\mbox{\boldmath{$ \sigma$}}^2 \cdot {\bf n}^2\right)\delta(t-t_1)\right)dt\right]}\nonumber\\
=\exp{\left[-i\left(\frac{A}{4}\left(\sum_{i=x,y,z}\sigma_i^1\sigma_i^2+1\right)\tau +\chi_1\mbox{\boldmath{$ \sigma$}}^1 \cdot {\bf n}^1 +\chi_2\mbox{\boldmath{$ \sigma$}}^2 \cdot {\bf n}^2\right)\right]}.\ \ \ \ \ \ \ \ \ \ \ \ \ \ \ \ \ \ \ \ \nonumber
\end{eqnarray}
If period of time $\tau$ tends to zero $\tau\rightarrow 0$ then the evolution operator $U_{II}$ takes the form
\begin{eqnarray}
&&U_{II}=e^{-i\left(\chi_1\mbox{\boldmath{$ \sigma$}}^1 \cdot {\bf n}^1 +\chi_2\mbox{\boldmath{$ \sigma$}}^2 \cdot {\bf n}^2\right)}\nonumber\\
&&=\left(\cos\chi_1-i\mbox{\boldmath{$ \sigma$}}^1 \cdot {\bf n}^1\sin\chi_1 \right)\left(\cos\chi_2-i\mbox{\boldmath{$ \sigma$}}^2 \cdot {\bf n}^2\sin\chi_2 \right).
\label{eq6}
\end{eqnarray}
Here we use the fact that $\mbox{\boldmath{$ \sigma$}}^1 \cdot {\bf n}^1$ and $\mbox{\boldmath{$ \sigma$}}^2 \cdot {\bf n}^2$ commute between themselves and $\left(\mbox{\boldmath{$ \sigma$}}^i \cdot {\bf n}^i\right)^2=1$.
The first and the second factors in this operator describe quantum evolution of the first and second spins under the external magnetic fields,
respectively.

Finally, if the evolution operator $U_{II}$ (\ref{eq6}) acts on the state $\vert\psi_I\rangle$ (\ref{eq4}) we obtain the following quantum state of two
qubits:
\begin{eqnarray}
&&\vert\psi\rangle=U_{II}\vert\psi_{I}\rangle=\cos\left(\frac{A}{2}t_1\right)\nonumber\\
&&\left[\left(\cos\chi_1-i\sin\chi_1\cos\theta_1\right)\vert\uparrow\rangle +\sin\chi_1 \sin\theta_1 e^{i\left(\phi_1 -\frac{\pi}{2}\right)}\vert\downarrow\rangle\right]\nonumber\\
&&\left[\sin\chi_2\sin\theta_2 e^{-i\left(\phi_2+\frac{\pi}{2}\right)}\vert\uparrow\rangle+\left(\cos\chi_2+i\sin\chi_2\cos\theta_2\right)\vert\downarrow\rangle\right]\nonumber\\
&&+\sin\left(\frac{A}{2}t_1\right)e^{-i\frac{\pi}{2}}\nonumber\\
&&\left[\sin\chi_1\sin\theta_1 e^{-i\left(\phi_1+\frac{\pi}{2}\right)}\vert\uparrow\rangle+\left(\cos\chi_1+i\sin\chi_1\cos\theta_1\right)\vert\downarrow\rangle\right]\nonumber\\
&&\left[\left(\cos\chi_2-i\sin\chi_2\cos\theta_2\right)\vert\uparrow\rangle +\sin\chi_2 \sin\theta_2 e^{i\left(\phi_2 -\frac{\pi}{2}\right)}\vert\downarrow\rangle\right].
\label{eq7}
\end{eqnarray}
This state is defined by seven real parameters. But in fact one parameter can be considered as an arbitrary phase of the state and can be omitted. So,
the quantum state (\ref{eq7}) contains only six independent parameters. For each predefined state of two qubits there exists a defined set of these parameters.

The state (\ref{eq7}) can be easily represented in the form of the Schmidt decomposition
\begin{eqnarray}
\vert\psi\rangle=\sum_{i=1,2}c_i\vert\alpha_i\rangle\vert\beta_i\rangle.
\label{eq12}
\end{eqnarray}
Single-particle states $\vert\alpha_i\rangle$ are related to the first spin take the form:
\begin{eqnarray}
&&\vert\alpha_1\rangle=\alpha\vert\uparrow\rangle +\alpha'\vert\downarrow\rangle,\nonumber\\
&&\vert\alpha_2\rangle=e^{-i\frac{\pi}{4}}\left(-{\alpha'}^*\vert\uparrow\rangle +{\alpha}^*\vert\downarrow\rangle\right)
\label{eq13}
\end{eqnarray}
and $\vert\beta_i\rangle$ are related to the second spin and they take the form:
\begin{eqnarray}
&&\vert\beta_1\rangle=\beta\vert\uparrow\rangle +\beta'\vert\downarrow\rangle,\nonumber\\
&&\vert\beta_2\rangle=e^{-i\frac{\pi}{4}}\left({\beta'}^*\vert\uparrow\rangle -{\beta}^*\vert\downarrow\rangle\right).
\label{eq14}
\end{eqnarray}
Here we introduce the following notation:
\begin{eqnarray}
&&c_1=\cos\left(\frac{A}{2}t_1\right),\quad c_2=\sin\left(\frac{A}{2}t_1\right),\label{eq1111}\\
&&\alpha=\cos\chi_1-i\sin\chi_1\cos\theta_1=\vert\alpha\vert e^{-i\gamma},\quad \alpha'=\sin\chi_1 \sin\theta_1 e^{i\left(\phi_1 -\frac{\pi}{2}\right)},\label{eq1112}\\
&&\beta=\sin\chi_2\sin\theta_2 e^{-i\left(\phi_2+\frac{\pi}{2}\right)},\quad \beta'=\cos\chi_2+i\sin\chi_2\cos\theta_2 =\vert\beta'\vert e^{i\eta},
\label{eq11}
\end{eqnarray}
where
\begin{eqnarray}
\vert\alpha\vert =\sqrt{\cos^2\chi_1+\sin^2\chi_1\cos^2\theta_1},\quad \tan\gamma =\tan\chi_1\cos\theta_1,\label{eq111}\\
\vert\beta'\vert =\sqrt{\cos^2\chi_2+\sin^2\chi_2\cos^2\theta_2},\quad \tan\eta =\tan\chi_2\cos\theta_2.
\label{eq112}
\end{eqnarray}
Note that the states belonging to the same spin are orthogonal
\begin{eqnarray}
\langle\alpha_i\vert\alpha_j\rangle=\delta_{ij},\quad \langle\beta_i\vert\beta_j\rangle=\delta_{ij},
\label{eq15}
\end{eqnarray}
and the normalization condition has the form
\begin{eqnarray}
{c_1}^2+{c_2}^2=1.
\label{eq16}
\end{eqnarray}

The single-particle states $\vert\alpha_1\rangle$ and $\vert\alpha_2\rangle$ (\ref{eq13}) are defined by three real independent parameters
$\vert\alpha\vert$, $\gamma$ and $\phi_1$. Remember that $\alpha$ and $\alpha'$ are connected with each other by the condition $\vert\alpha\vert^2+\vert\alpha'\vert^2=1$.
It is easy to see that the $\phi_1$ takes any values $[0,2\pi]$ regardless of the
other parameters. Let us show that $\vert\alpha\vert$ and $\gamma$ are independent parameters.
From the second equations in (\ref{eq1112}) and (\ref{eq111}) we obtain
\begin{eqnarray}
\tan\gamma=\sqrt{1-\frac{\vert\alpha'\vert ^2}{\sin^2\chi_1}}\tan\chi_1.
\label{eq163}
\end{eqnarray}
The necessary condition to satisfy this equation can be written as $\sin^2\chi_1 \geq\vert\alpha'\vert^2$. From (\ref{eq163}) follows that
$\gamma$ takes any value $-\infty<\tan\gamma<+\infty$ regardless of $\vert\alpha'\vert$.
This proves that this parameters are independent. A similar situation is in the case with $\vert\beta_1\rangle$ and $\vert\beta_2\rangle$
(\ref{eq14}), which also contain three real independent parameters.

{\it Example}. Let us find the conditions for the creation
of the unpolarized triplet state (\ref{eq2(1)}), which is an eigenvector of the Heisenberg Hamiltonian (\ref{eq1}). The comparison of (\ref{eq7})
with (\ref{eq2(1)}) leads to the following set of parameters:
\newline
1. $A > 0$, $t_1=\frac{\pi}{2 A}$, $\theta_1=0$, $\theta_2=0$ and $\chi_1-\chi_2=\frac{\pi}{4}$
\newline
or
\newline
2. $A < 0$, $t_1=\frac{\pi}{2\vert A\vert}$, $\theta_1=0$, $\theta_2=0$ and $\chi_1-\chi_2=-\frac{\pi}{4}$.
\newline
Let us explain step by step how the unpolarized triplet state can be prepared. At the first step two spins interact during the period of
time $t_1=\frac{\pi}{2\vert A\vert}$ as (\ref{eq1}) and from the initial state $\vert\uparrow\downarrow\rangle$ we obtain the following
$\frac{1}{\sqrt{2}}\left(\vert\uparrow\downarrow\rangle-i\vert\downarrow\uparrow\rangle\right)$ ($A > 0$) or
$\frac{1}{\sqrt{2}}\left(\vert\uparrow\downarrow\rangle+i\vert\downarrow\uparrow\rangle\right)$ ($A < 0$) state.
At the second step, at the moment of time $t_1$ we apply pulsed magnetic fields individually to each spin directed along the $z$-axis
($\theta_1=\theta_2=0$).
The difference between the values of these fields are $\chi_1-\chi_2=\frac{\pi}{4}$ ($A > 0$) or $\chi_1-\chi_2=-\frac{\pi}{4}$ ($A < 0$).
The final state equals to the unpolarized triplet state (\ref{eq2(1)}) modulo a global phase $-\frac{\pi}{4}$ in the case of $A > 0$ and $\frac{\pi}{4}$
in the case of $A < 0$.

\section{Preparation of two spin-$\frac{1}{2}$ states in physically realizable systems \label{sec3}}

We propose a method of creation of the quantum state using a physical system of the atom which has a nuclear spin $I=1/2$ and a valence electron
(with the spin $S=1/2$). Modern experimental technics do not allow to control individually each
spin with help of the pulsed magnetic field considered at the second step of the two-steps method in Sec. \ref{sec2}. Therefore, we propose to examine a
simplified version of this method, which is physically applicable.
Today, experimental techniques allow to control the evolution of the electron spin using the magnetic field of about
mT \cite{esr}. It means that value of the interaction between the electron spin and the magnetic field is approximately equals to the value
of the hyperfine interaction $A$. Then, the Hamiltonian, which describes the system of the atom in magnetic field, contains the components that
are responsible for
description of the interaction of the system with the magnetic field $B$ and the interaction between the electron and the nuclear spins
\begin{eqnarray}
H=\gamma_e B {\bf S} \cdot {\bf n}-\gamma_n B{\bf I} \cdot {\bf n} +A{\bf S}\cdot {\bf I},
\label{eq17}
\end{eqnarray}
where ${\bf n}=(\sin\theta\cos\phi, \sin\theta\sin\phi, \cos\theta)$ is a direction of the magnetic field, $\theta$ and $\phi$ are spherical angles,
$\gamma_e$ (or $\gamma_n$) is the gyromagnetic ratio of the electron (or nucleus).
The Hamiltontian (\ref{eq17}) does not allow to prepare arbitrary quantum state of two qubits because it does not contain the sufficient number of
parameters.

The Hamiltonian (\ref{eq17}) has four energy levels
\begin{eqnarray}
&&E_1=\frac{\gamma_e B}{2}-\frac{\gamma_n B}{2} + \frac{A}{4},\nonumber\\
&&E_2=\frac{1}{2}\sqrt{\left(\gamma_e B+\gamma_n B\right)^2 +A^2} - \frac{A}{4},\nonumber\\
&&E_3=-\frac{1}{2}\sqrt{\left(\gamma_e B+\gamma_n B\right)^2 +A^2} - \frac{A}{4},\nonumber\\
&&E_4=-\frac{\gamma_e B}{2}+\frac{\gamma_n B}{2} + \frac{A}{4}
\label{eq18}
\end{eqnarray}
with the corresponding eigenvectors
\begin{eqnarray}
&&\vert\psi_1\rangle=\vert + +\rangle,\nonumber\\
&&\vert\psi_2\rangle=\cos\frac{\eta}{2}\vert + -\rangle +\sin\frac{\eta}{2}\vert - +\rangle,\nonumber\\
&&\vert\psi_3\rangle=-\sin\frac{\eta}{2}\vert + -\rangle +\cos\frac{\eta}{2}\vert - +\rangle,\nonumber\\
&&\vert\psi_4\rangle=\vert - -\rangle,
\label{eq19}
\end{eqnarray}
where $\vert +\rangle =\cos\frac{\theta}{2}\vert\uparrow\rangle+\sin\frac{\theta}{2}e^{i\phi}\vert\downarrow\rangle$ and
$\vert -\rangle =-\sin\frac{\theta}{2}\vert\uparrow\rangle+\cos\frac{\theta}{2}e^{i\phi}\vert\downarrow\rangle$,
$\tan \eta =\frac{A}{\gamma_e B+\gamma_n B}$. Here, $\eta$ is the angle between the direction of external
magnetic field and actual electron and nuclear spins precession axis \cite{data3}.
The left (right) ket represents the electron (nuclear) spin state.

The initial state we put as $\vert + -\rangle$ because this allows us to control the system of two spins using the
magnetic field directed along the $z$-axis and to simplify the calculations. The initial state can be prepared if the system of two spins is
placed in strong magnetic field ($B\gg A$). Eigenvalues and eigenvectors of this system approximately are
\begin{eqnarray}
&&\vert + +\rangle:\quad \frac{\gamma_e B}{2}-\frac{\gamma_n B}{2} + \frac{A}{4},\nonumber\\
&&\vert + -\rangle:\quad \frac{\gamma_e B}{2}+\frac{\gamma_n B}{2} - \frac{A}{4},\nonumber\\
&&\vert - +\rangle:\quad -\frac{\gamma_e B}{2}-\frac{\gamma_n B}{2} - \frac{A}{4},\nonumber\\
&&\vert - -\rangle:\quad -\frac{\gamma_e B}{2}+\frac{\gamma_n B}{2} + \frac{A}{4}
\label{eq18(1)}
\end{eqnarray}
because $\cos\frac{\eta}{2}\simeq 1$ and $\sin\frac{\eta}{2}\simeq 0$  \cite{phosphorus1, phosphorus2, phosphorus5}.
The electron spin resonance technique easily allows to prepare factorized initial state
$\vert + -\rangle$ similarly as it was made in \cite{phosphorus1}.

Now, using the magnetic field $B_z$ which is directed along the $z$-axis we can start preparation of a quantum state on the atom system. The Hamiltonian
(\ref{eq17}) can be rewritten as follows:
\begin{eqnarray}
H=H_{xy}+H_{zz}+H_{+},
\label{eq21}
\end{eqnarray}
where
\begin{eqnarray}
&&H_{xy}=\omega_-\left(S_z-I_z\right)+A\left(S_xI_x+S_yI_y\right),\label{eq22}\\
&&H_{zz}=AS_zI_z,\label{eq24}\\
&&H_+=\omega_+\left(S_z+I_z\right).\label{eq23}
\end{eqnarray}
Here we denoted $\omega_-=\frac{\gamma_eB_z+\gamma_nB_z}{2}$ and $\omega_+=\frac{\gamma_eB_z-\gamma_nB_z}{2}$. Operators $H_{xy}$, $H_+$ and $H_{zz}$ commute
between themselves. Note that we
do not take into account effects which distort the square shape of a pulsed magnetic field.
The evolution operator for this system (\ref{eq21}) takes the form
\begin{eqnarray}
U(t)=e^{-iH_{xy}t}e^{-iH_{zz}t}e^{-iH_+t},
\label{eq25}
\end{eqnarray}
where
\begin{eqnarray}
&&e^{-iH_{xy}t}=1+\left(\cos\left(\Omega t\right)-1\right)\left(\frac{1}{2}-2S_zI_z\right)-i\frac{\sin\left(\Omega t\right)}{\Omega}H_{xy},\label{eq26}\\
&&e^{-iH_{zz}t}=\cos\left(\frac{A}{4}t\right)-i\frac{4\sin\left(\frac{A}{4}t\right)}{A}H_{zz},\label{eq28}\\
&&e^{-iH_+t}=1+\left(\cos\left(\omega_+ t\right)-1\right)\left(\frac{1}{2}+2S_zI_z\right)-i\frac{\sin\left(\omega_+ t\right)}{\omega_+}H_+.\label{eq27}
\end{eqnarray}
Here we use that $H_{xy}^{2n}=\Omega^{2n}\left(\frac{1}{2}-2S_zI_z\right)$, $H_{xy}^{2n+1}=\Omega^{2n}H_{xy}$, $H_{zz}^2=\left(\frac{A}{4}\right)^2$,
$H_+^{2n}=\omega_+^{2n}\left(\frac{1}{2}+2S_zI_z\right)$ and $H_+^{2n+1}=\omega_+^{2n}H_+$,
where $\Omega=\sqrt{{\omega_-}^2+\left(\frac{A}{2}\right)^2}$ and $n=1,2,3,\ldots$
In the basis labeled as $\vert\uparrow\uparrow\rangle$, $\vert\uparrow\downarrow\rangle$, $\vert\downarrow\uparrow\rangle$
and $\vert\downarrow\downarrow\rangle$, the evolution operator $U(t)$ can be represented as:
\begin{eqnarray}
U(t)=U_1U_2,
\label{eq25(matrix)}
\end{eqnarray}
where
\begin{eqnarray}
U_1=\left( \begin{array}{ccccc}
e^{-i\left(\omega_++\frac{A}{4}\right)t} & 0 & 0 & 0 \\
0 & 1 & 0 & 0 \\
0 & 0 & 1 & 0 \\
0 & 0 & 0 & e^{i\left(\omega_+-\frac{A}{4}\right)t}
\end{array}\right)
\label{eq25(matrix1)}
\end{eqnarray}
and
\begin{eqnarray}
U_2=\left( \begin{array}{ccccc}
1 & 0 & 0 & 0 \\
0 & \left(\cos\left(\Omega t\right)-i\frac{\omega_-}{\Omega}\sin\left(\Omega t\right)\right)e^{i\frac{A}{4}t} & -i\frac{A}{2\Omega}\sin\left(\Omega t\right)e^{i\frac{A}{4}t} & 0 \\
0 & -i\frac{A}{2\Omega}\sin\left(\Omega t\right)e^{i\frac{A}{4}t} & \left(\cos\left(\Omega t\right)+i\frac{\omega_-}{\Omega}\sin\left(\Omega t\right)\right)e^{i\frac{A}{4}t} & 0 \\
0 & 0 & 0 & 1
\end{array}\right).
\label{eq25(matrix2)}
\end{eqnarray}

Using (\ref{eq25})--(\ref{eq27}) we obtain
\begin{eqnarray}
&&\vert\psi\rangle =U(t)\vert + -\rangle = e^{-i\omega_+t}e^{-i\frac{A}{4}t}e^{i\pi}\cos\frac{\theta}{2}\sin\frac{\theta}{2}\vert\uparrow\uparrow\rangle\nonumber\\
&&+e^{i\frac{A}{4}t}e^{i\phi}\left[\cos\left(\Omega t\right)\cos^2\frac{\theta}{2}+i\left(\frac{A}{2\Omega}\sin^2\frac{\theta}{2}-\frac{\omega_-}{\Omega}\cos^2\frac{\theta}{2}\right)\sin\left(\Omega t\right)\right]\vert\uparrow\downarrow\rangle\nonumber\\
&&+e^{i\frac{A}{4}t}e^{i\phi}e^{i\pi}\left[\cos\left(\Omega t\right)\sin^2\frac{\theta}{2}+i\left(\frac{A}{2\Omega}\cos^2\frac{\theta}{2}+\frac{\omega_-}{\Omega}\sin^2\frac{\theta}{2}\right)\sin\left(\Omega t\right)\right]\vert\downarrow\uparrow\rangle\nonumber\\
&&+e^{i\omega_+t}e^{-i\frac{A}{4}t}e^{2i\phi}\cos\frac{\theta}{2}\sin\frac{\theta}{2}\vert\downarrow\downarrow\rangle.
\label{eq29}
\end{eqnarray}
From the analysis of this state it is clear that it contains an arbitrary state on the subspace spanned by $\vert\uparrow\downarrow\rangle$,
$\vert\downarrow\uparrow\rangle$ ($\theta=0$)
\begin{eqnarray}
\vert\psi\rangle=\left(\cos\left(\Omega t\right)-i\frac{\omega_-}{\Omega}\sin\left(\Omega t\right)\right)\vert\uparrow\downarrow\rangle +\frac{A}{2\Omega}\sin\left(\Omega t\right)e^{-i\frac{\pi}{2}}\vert\downarrow\uparrow\rangle.
\label{eq30}
\end{eqnarray}
In other words, state (\ref{eq30}) can be reached starting from the initial state $\vert\uparrow\downarrow\rangle$.
In turn, state (\ref{eq30}) contains the maximally entangled state
\begin{eqnarray}
\vert\psi_{{\rm ENT}}\rangle=\frac{1}{\sqrt{2}}\left(\vert\uparrow\downarrow\rangle+e^{i\chi}\vert\downarrow\uparrow\rangle\right),
\label{eq31}
\end{eqnarray}
when the following conditions are satisfied
\begin{eqnarray}
\cot\chi =-\frac{\omega_-}{\sqrt{\Omega^2-2\omega_-^2}},\quad \tan^2\left(\Omega t\right)=\frac{\Omega^2}{\Omega^2-2\omega_-^2},
\label{eq33}
\end{eqnarray}
where $\chi$ is an arbitrary phase.

From the first equation in (\ref{eq33}) we find the value of the magnetic field
\begin{eqnarray}
B_z=\frac{A\cos\chi}{\gamma_e+\gamma_n}
\label{eq34}
\end{eqnarray}
which allows us to reach the maximally entangled state ({\ref{eq31}}) during the period of time
\begin{eqnarray}
t=\frac{2}{A\sqrt{1+\cos^2\chi}}\arctan\sqrt{1+2\cot^2\chi}.
\label{eq35}
\end{eqnarray}
This equation is obtained from the second equation of (\ref{eq33}) using equation (\ref{eq34}).

\section{Application to the $^{31}$P system \label{sec4}}

The $^{31}$P donor in silicon has the nuclear spin $I=1/2$, and at cryogenic temperatures an electron can be caught by this atom.
Therefore, a single $^{31}$P donor can be considered as a two-qubit system. Interaction of the $^{31}$P system with the external
magnetic field is proportional to their gyromagnetic ratios: $\gamma_n=17.23$ MHz T$^{-1}$ \cite{data2} for the nucleus and $\gamma_e=27.97$ GHz T$^{-1}$
for the electron. Interaction between the nucleus and the electron is defined by the hyperfine interaction $A=117.53$ MHz \cite{data1}.
Long spin coherence time \cite{phosphorus2, phosphorus6, phosphorus8} allows us to manipulate this system successfully \cite{phosphorus2}.

Unpolarized triplet state (\ref{eq2(1)}) can be easily prepared using the phosphorus donor in silicon starting from the
$\vert\uparrow\downarrow\rangle$ state, which is an eigenstate of this system in the strong magnetic field $B_z\gg A$.
The creation of an unpolarized triplet state is important because this is the maximally entangled state.
The condition for the preparation of the state (\ref{eq2(1)}) follows from the equation (\ref{eq31}): $\chi=0$.
If we put $\chi=0$ in the relations (\ref{eq34}) and (\ref{eq35}) we obtain necessary conditions for preparation the unpolarized triplet state: the value of
the external magnetic field $B_z=\frac{A}{\gamma_e+\gamma_n}\simeq 4.2$ mT
and the period of time of the evolution $t=\frac{\pi}{\sqrt{2}A}\simeq 19$ ns. To reach the unpolarized triplet state from $\vert\uparrow\downarrow\rangle$,
the system of the phosphorus donor in silicon is placed in the
magnetic field $B_z\simeq 4.2$ mT during the period of time of $t\simeq 19$ ns. The period of time of the evolution is much shorter than the coherence lifetime $2$ s of
the electron spin of the phosphorus donor in silicon \cite{phosphorus6}. Short pulse of the magnetic fields are considered
in different areas (see for instance \cite{PMF1, PMF2, PMF3}). The possibility of application of short pulse of the magnetic field on experimental level
in our case can be a subject of following studies.

\begin{figure*}[h]
	\centering
		\includegraphics[width=0.5\textwidth]{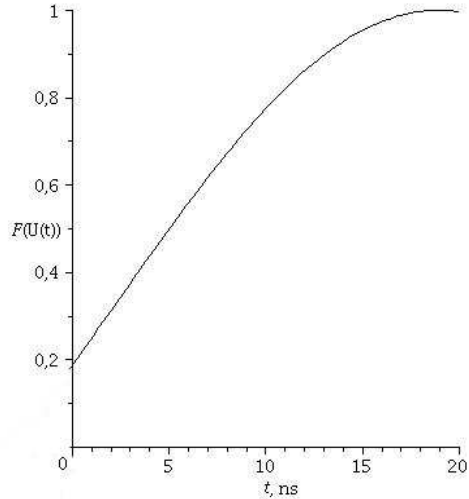}
	\caption{\textit{The fidelity between the evolution operator $U(t)$ (\ref{eq25}) and quantum gate $W$ (\ref{eq37}).}}
	\label{fidelity}
\end{figure*}

Here, we make a prediction regarding the fidelity with which the quantum gate, that allows to reach an unpolarized triplet state, can be done.
These calculations might be useful to experimentalists.
The fidelity between two unitary operators for the system of $N$ discrete levels may be represented as a scalar cost function \cite{FIDELITY}
\begin{eqnarray}
F(U(t))=\frac{1}{N}\Re(Tr[W^+U(t)]),
\label{eq36}
\end{eqnarray}
where $W$ is the quantum gate which is obtained from (\ref{eq25})-(\ref{eq27}) or (\ref{eq25(matrix)})-(\ref{eq25(matrix2)}) using that $B_z=\frac{A}{\gamma_e+\gamma_n}$ and
$t=\frac{\pi}{\sqrt{2}A}$. In the matrix representation quantum gate $W$ reads:
\begin{eqnarray}
W=\left( \begin{array}{ccccc}
e^{-i\frac{\pi}{2\sqrt{2}}\left(\frac{\gamma_e-\gamma_n}{\gamma_e+\gamma_n}+\frac{1}{2}\right)} & 0 & 0 & 0 \\
0 & -\frac{i}{\sqrt{2}}e^{i\frac{\pi}{4\sqrt{2}}} & -\frac{i}{\sqrt{2}}e^{i\frac{\pi}{4\sqrt{2}}} & 0 \\
0 & -\frac{i}{\sqrt{2}}e^{i\frac{\pi}{4\sqrt{2}}} & \frac{i}{\sqrt{2}}e^{i\frac{\pi}{4\sqrt{2}}} & 0 \\
0 & 0 & 0 & e^{i\frac{\pi}{2\sqrt{2}}\left(\frac{\gamma_e-\gamma_n}{\gamma_e+\gamma_n}-\frac{1}{2}\right)}
\end{array}\right).
\label{eq37}
\end{eqnarray}
If we insert (\ref{eq25(matrix)})-(\ref{eq25(matrix2)}) and (\ref{eq37}) in (\ref{eq36}) and using that $N=4$ we obtain the expression for fidelity
in the following form:
\begin{eqnarray}
F(U(t))=\frac{1}{2}\cos\left(\frac{A}{4}t-\frac{\pi}{4\sqrt{2}}\right)\left(\cos\left(\frac{\gamma_e-\gamma_n}{\gamma_e+\gamma_n}\left(\frac{At}{2}-\frac{\pi}{2\sqrt{2}}\right)\right)+\sin\left(\frac{A}{\sqrt{2}}t\right)\right).
\label{eq38}
\end{eqnarray}
The graph of the function (\ref{eq38}) is shown  in Fig. \ref{fidelity}.

\section{Conclusion \label{sec5}}

The two-step method suggested in this paper allows to prepare an arbitrary quantum state of a two-qubit system.
This method naturally allows to prepare the final state in the form of the Schmidt decomposition.
At the first step, the quantum state driven by an isotropic Heisenberg Hamiltonian evolves from the initial state $\vert\uparrow\downarrow\rangle$
to some state which depends on the period of time of the evolution $t_1$. Note that we do not start with $\vert\uparrow\uparrow\rangle$ or $\vert\downarrow\downarrow\rangle$
states because these states are eigenstates of this system. Then at the second step, at the moment of time $t_1$ we apply individually to
each spin the pulsed magnetic fields (\ref{eq5}) and obtain the final quantum state (\ref{eq7}). This state is defined by six independent parameters:
period of time $t_1$, and parameters which determine the magnetic field for the first and the second spin. For each predefined state of two qubits there exists a
defined set of these parameters. We hope that the proposed method can be realized in future because modern experimental technologies do not allow to control
individually each spin using the pulsed magnetic field which is used at the second step of the method. Therefore, we propose to consider a simplified version
of this method, which is physically applicable. The experimental technique allows to control the evolution of the spins using the uniform
magnetic field with $B\gamma_s$ up to $10^3$ MHz (where $\gamma_s$ is the gyromagnetic ratio of the spin) (for example, see \cite{esr}) corresponding to the
value of interaction between the nuclear spin $I=1/2$ and the valence electron in an atom. Therefore, we consider the evolution of such a system
in an external uniform magnetic field oriented along a certain direction. The evolution of this system is determined by four parameters: period of time of the
evolution, value of the magnetic field and two angles that define the direction of the field. This approach does not allow to prepare
an arbitrary quantum state because the number of parameters determining the evolution of the system is not sufficient. Although, we
can create an arbitrary state on the subspace spanned by $\vert\uparrow\downarrow\rangle$, $\vert\downarrow\uparrow\rangle$ (\ref{eq30}).

The conditions for the preparation of the unpolarized triplet state (\ref{eq2(1)}) using the physical system of $^{31}$P were found. Namely, we have shown
that the unpolarized triplet state can be reached starting from the initial state $\vert\uparrow\downarrow\rangle$ if the magnetic field of about $4.2$ mT
is applied along the $z$-axis during the period of time of $19$ ns.

\section{Acknowledgment}

The authors would like to thank Dr. A. Rovenchak and Dr M Stetsko for useful comments.


\begin{thebibliography}{99}
\bibitem{quantcomp} David P. DiVincenzo, Fortschr. Phys. \textbf{48}, 771 (2000).
\bibitem{qdots1} Daniel Loss and David P. DiVincenzo, Phys. Rev. A \textbf{57}, 120 (1998).
\bibitem{qdots2} Guido Burkard, Daniel Loss. and David P. DiVincenzo, Phys. Rev. B \textbf{59}, 2070 (1999).
\bibitem{qdots3} Niklas Rohling and Guido Burkard, Phys. Rev. B \textbf{88}, 085402 (2013).
\bibitem{scc} J. R. Petta, A. C. Johnson, J. M. Taylor, E. A. Laird, A. Yacoby, M. D. Lukin, C. M. Marcus, M. P. Hanson and A.C. Gossard,
Science \textbf{309}, 2180 (2005).
\bibitem{semcond1} S. Kuhlen, K. Schmalbuch, M. Hagedorn, P. Schlammes, M. Patt, M. Lepsa, G. G\"untherodt and B. Beschoten, Phys. Rev. Lett. \textbf{109},
146603 (2012).
\bibitem{semcond2} Rutger Vrijen, Eli Yablonovitch, Kang Wang, Hong Wen Jiang, Alex Balandin, Vwani Roychowdhury, Tal Mor and David DiVincenzo,
Phys. Rev. A \textbf{62}, 012306 (2000).
\bibitem{semcond3} Hendrik Bluhm, Sandra Foletti, Izhar Neder, Mark Rudner, Diana Mahalu, Vladimir Umansky and Amir Yacoby, Nature Physics \textbf{7}, 109 (2011).
\bibitem{semcond4} F. H. L. Koppens, C. Buizert, K. J. Tielrooij, I. T. Vink, K. C. Nowack, T. Meunier, L. P. Kouwenhoven and L. M. K. Vandersypen,
Nature \textbf{442}, 766 (2006).
\bibitem{semcond5} K. C. Nowack, F. H. L. Koppens, Yu. V. Nazarov and L. M. K. Vandersypen, Science \textbf{318}, 1430 (2007).
\bibitem{semcond6} B. M. Maune et al., Nature \textbf{481}, 344 (2012).
\bibitem{supcond1} L. F. Wei, Yu-xi Liu and Franco Nori, Phys. Rev. B \textbf{71}, 134506 (2005).
\bibitem{supcond2} J. E. Mooij, T. P. Orlando, L. Levitov, Lin Tian, Caspar H. van der Wal  and Seth Lloyd, Science \textbf{285}, 1036 (1999).
\bibitem{supcond3} Yuriy Makhlin, Gerd Sch\"on and Alexander Shnirman, Rev. Mod. Phys. \textbf{73}, 357 (2001).
\bibitem{supcond4} J. Majer et al., Nature \textbf{449}, 443 (2007).
\bibitem{phosphorus3} B. E. Kane, Nature \textbf{393}, 133 (1998).
\bibitem{phosphorus1} Jarryd J. Pla, Kuan Y. Tan, Juan P. Dehollain, Wee H. Lim, John J. L. Morton, Floris A. Zwanenburg, David N. Jamieson,
Andrew S. Dzurak and Andrea Morello, Nature \textbf{496}, 334 (2013).
\bibitem{phosphorus2} Jarryd J. Pla, Kuan Y. Tan, Juan P. Dehollain, Wee H. Lim, John J. L. Morton, David N. Jamieson,
Andrew S. Dzurak and Andrea Morello, Nature \textbf{489}, 541 (2012).
\bibitem{phosphorus4} Andrea Morello et al., Nature \textbf{467}, 687 (2010).
\bibitem{phosphorus6} Alexei M. Tyryshkin et al., Nature Materials \textbf{11}, 143 (2012).
\bibitem{phosphorus7} John J. L. Morton, Alexei M. Tyryshkin, Richard M. Brown, Shyam Shankar, Brendon W. Lovett, Arzhang Ardavan, Thomas Schenkel,
Eugene E. Haller, Joel W. Ager and S. A. Lyon, Nature \textbf{455}, 1085 (2008).
\bibitem{phosphorus8} M. Steger, K. Saeedi, M. L. W. Thewalt, J. J. L. Morton, H. Riemann, N. V. Abrosimov, P. Becker and H. J. Pohl,
Science \textbf{336}, 1280 (2012).
\bibitem{srtech} C. P. Slichter, {\it Principles of Magnetic Resonance} (Springer-Verlag, Berlin, 1990).
\bibitem{sq1} R. Hanson, L. P. Kouwenhoven, J. R. Petta, S. Tarucha and L. M. K. Vandersypen, Rev. Mod. Phys. \textbf{79}, 1217 (2007).
\bibitem{sq2} E. A. Laird, J. R. Petta, A. C. Johnson, C. M. Marcus, A. Yacoby, M. P. Hanson and A. C. Gossard, Phys. Rev. Lett. \textbf{97}, 056801 (2006).
\bibitem{sq3} John J. L. Morton, Dane R. McCamey, Mark A. Eriksson and Stephen A. Lyon, Nature. \textbf{479}, 345 (2011).
\bibitem{sq4} Iulia Buluta, Sahel Ashhab and Franco Nori, Rep. Prog. Phys. \textbf{74}, 104401 (2011).
\bibitem{BiinSi} M. H. Mohammady, G. W. Morley, A. Nazir and T. S. Monteiro, Phys. Rev. B \textbf{85}, 094404 (2012).
\bibitem{HinSi} A. J. Skinner, M. E. Davenport and B. E. Kane, Phys. Rev. Lett. \textbf{90}, 087901 (2003).
\bibitem{LNT} Kamyar Saeedi, Stephanie Simmons, Jeff Z. Salvail, Phillip Dluhy, Helge Riemann, Nikolai V. Abrosimov, Peter Becker, Hans-Joachim Pohl,
John J. L. Morton and Mike L. W. Thewalt, Science \textbf{342}, 830 (2013).
\bibitem{phosphorus9} Andre R. Stegner, Christoph Boehme, Hans Huebl, Martin Stutzmann, Klaus Lips and Martin S. Brandt, Nat. Phys. \textbf{2}, 835 (2006).
\bibitem{phosphorus10} Hans Huebl, Felix Hoehne, Benno Grolik, Andre R. Stegner, Martin Stutzmann and Martin S. Brandt,
Phys. Rev. Lett. \textbf{100}, 177602 (2008).
\bibitem{phosphorus11} G. W. Morley, D. R. McCamey, H. A. Seipel, L. C. Brunel, J. van Tol and C. Boehme, Phys. Rev. Lett. \textbf{101}, 207602 (2008).
\bibitem{phosphorus12} Felix Hoehne, Hans Huebl, Bastian Galler, Martin Stutzmann and Martin S. Brandt, Phys. Rev. Lett. \textbf{104}, 046402 (2010).
\bibitem{CRSP} Xiu-Bo Chen, Song-Ya Ma, Yuan Su, Ru Zhang and Yi-Xian Yang, Quantum Inf Process \textbf{11}, 1653 (2012).
\bibitem{GHZP} Kui Hou, Yi-Bao Li, Gou-Hong Liu and Shou-Qi Sheng, J. Phys. A \textbf{44}, 255304 (2011).
\bibitem{brach} A. R. Kuzmak, V. M. Tkachuk, J. Phys. A \textbf{46}, 155305 (2013).
\bibitem{esr} H. Sanada, Y. Kunihashi, H. Gotoh, K. Onomitsu, M. Kohda, J. Nitta, P. V. Santos and T. Sogawa, Nature Physics \textbf{9}, 280 (2013).
\bibitem{data3} A. Schweiger and G. Jeschke, {\it Principles of Pulse Electron Paramagnetic Resonance} (Oxford, 2001), Chap. 3, Sec. 3.5, pp.58-62.
\bibitem{phosphorus5} H. Morishita, L. S. Vlasenko, H. Tanaka, K. Semba, K. Sawano, Y. Shiraki, M. Eto and K. M. Itoh, Phys. Rev. B. \textbf{80}, 205206 (2009).
\bibitem{data2} M. Steger et al., J. Appl. Phys. \textbf{109}, 102411 (2011).
\bibitem{data1} G. Feher, Phys. Rev. \textbf{114}, 1219 (1959).
\bibitem{PMF1} Th. Gerrits, H. A. M. van den Berg, J. Hohlfeld, L. B\"ar and Th. Rasing, Nature {\bf 418}, 509 (2002).
\bibitem{PMF2} P. Jehenson, M. Westphal, N. Schuff, J. Mag. Res.{\bf 90}, 264 (1969).
\bibitem{PMF3} M. Bauer, R. Lopusnik, J. Fassbender and B. Hillebrands, Appl. Phys. Lett. 76, 2758 (2000).
\bibitem{FIDELITY} Renan Cabrera, Ofer M. Shir, Rebing Wu and Herschel Rabitz, J. Phys. A \textbf{44}, 095302 (2011).
\end{thebibliography}
\end{document}